\documentclass[12pt,preprint]{aastex}
\shorttitle{ON THE BINARY NATURE OF J1628}
\shortauthors{TORRES ET AL.}
\begin{document}

\title{On the binary nature of 1RXS J162848.1-415241}  
\author{M. A. P. Torres\altaffilmark{1}, M. R. Garcia\altaffilmark{1}, D. Steeghs\altaffilmark{1}, J. E. McClintock\altaffilmark{1}}

\altaffiltext{1}{Harvard-Smithsonian Center for Astrophysics, 60 Garden St, Cambridge, MA 02138; mtorres@cfa.harvard.edu,mgarcia@cfa.harvard.edu,dsteeghs@cfa.harvard.edu,jem@cfa.harvard.edu}

\begin{abstract}
We present spectroscopy of the optical counterpart to 1RXS
J162848.1-41524, also known as the microquasar candidate MCQC
J162847-4152. All the data indicate that this X-ray source is not a
microquasar, and that it is a single-lined chromospherically active
binary system with a likely orbital period of 4.9 days. Our analysis
supports a K3IV spectral classification for the star, which is dominant at
optical wavelengths. The unseen binary component is most likely
a late-type (K7-M) dwarf or a white dwarf. Using the high resolution
spectra we have measured the K3 star's rotational broadening to
be $v \sin i = 43 \pm 3$ km s$^{-1}$ and determined a lower limit to
the binary mass ratio of $q(={M_2}/{M_1})>2.0$. The high rotational
broadening together with the strong Ca {\sc ii} H \& K/H$\alpha$
emission and high-amplitude photometric variations indicate that the
evolved star is very chromospherically active and responsible for the
X-ray/radio emission.
\end{abstract}


\section{INTRODUCTION}

X-ray binaries are binary systems where a neutron star or a black
hole accretes matter from its companion star. Of $\sim280$ known X-ray
binaries, fifteen show persistent or episodic relativistic radio jets
and are called microquasars as they are reminiscent of AGN (see
e.g. Paredes 2005 for an inventory and Mirabel \& Rodr\'\i guez 1999,
Fender 2004a for a review of their properties). Microquasars are of
great interest because they may provide a unique opportunity to gain
insight into the mechanisms ruling the formation and evolution of
relativistic outflows coupled to accretion phenomena. In this regard,
the variations observed in microquasars are perceptible on human
timescales, which allows studies impossible to perform for the more
plentiful (but slowly varying) extragalactic relativistic jet
sources. The sample of currently known microquasars is however small,
limiting any robust conclusions of how the jet formation and its
properties depend on parameters such as the accretion rate, accretion
geometry or the mass and spin of the compact object (see e.g. Garcia
et al. 2003). In an attempt to enlarge the sample, two long-term
programs have been carried out to identify microquasar candidates in
the Galaxy by cross-correlating the ROSAT All-Sky Survey (RASS) Bright
Source catalogue (Voges et al. 1999) with the NVSS, GB6 and/or PMN
radio surveys (Condon et al. 1998, Gregory et al. 1996, Griffith et
al. 1994). Both programs share similar selection criteria:
positionally coincident X-ray and radio sources that are not extended
and that have hard X-ray spectra typical of X-ray binaries (see
Tsarevsky et al. 2002 and Paredes, Rib\'o \& Mart\'\i~2002 for more
details). Once the optical counterpart is identified, spectroscopic
observations are carried out to establish the nature of the source,
and further follow-up observations are initiated to confirm its
candidacy.

1RXS J162848.1-415241 (hereafter J1628) was detected with the Position
Sensitive Proportional Counter (PSPC) during the ROSAT All-Sky Survey
at a count rate of 0.119 c s$^{-1}$ (0.1-2.4 keV). This X-ray
source was reported to be the best microquasar candidate found during
one of the surveys for new microquasars in the Galactic plane
(Tsarevsky et al. 2001, 2002). Tsarevsky et al. selected this ROSAT
object following the criteria outlined above, i.e: J1628 is a point-like
X-ray source with an X-ray spectral hardness typical of X-ray
binaries (Motch et al. 1998) and it is associated with a variable radio
source (Tsarevsky et al. 2001; Rupen, Mioduszewski \& Dhawan 2002,
2004; Slee et al. 2002). The optical counterpart to J1628 was
identified with the apparently variable stellar-like source GSC
07861-01088 (V=13.4). Low resolution optical spectroscopy showed a
K-type spectrum with a variable H$\alpha$ emission profile (Tsarevsky
et al. 2001). The photometric variability of J1628 was confirmed by
Buxton et al. (2004) who observed a 0.30 mag amplitude variation in
the V-band light curve. On the basis of all the gathered photometry,
they found a periodicity of $4.9364 \pm 0.0018$~d (Tsarevsky; private
communication). Buxton et al. interpreted the photometric variations
as arising from a double-humped ellipsoidal light curve (as is typical
in X-ray novae in quiescence) or from a pulsating star.

During the 2002-2004 seasons, J1628 was observed sporadically during
the spectroscopic programs conducted by several CfA astronomers
observing at Las Campanas Observatory (LCO). The aim of these
observations was to confirm the suggested microquasar nature of J1628
by deriving its binary parameters from the analysis of the
spectroscopic data accrued during these three years. The spectra
obtained during 2002 showed clear radial velocity variations from
night to night. 

The binary nature of J1628 was confirmed one year later upon the
acquisition of echelle spectra which showed highly broadened
photospheric lines due to the stellar rotation: it is well-known that
rapid rotation in G and K-type stars is a sign of a close binary where
tidal effects have synchronized stellar rotation and orbital
revolution. Of course, FK Comae stars (Bopp \& Stencel 1981) and young
single stars that have not yet been spun down by magnetic braking are
also fast rotators; however, they do not show radial velocity
variations. A pulsating nature for J1628 is ruled out because the only
Population I pulsating stars with low-amplitude ($\Delta V < 1$~mag)
light curves are Classical Cepheids with long periods; although these
stars do show a K-type spectra when they reach luminosity minimum
(Efremov 1975, Petit 1985), they are known to be slow rotators (Kraft
1966).

In this paper we make a detailed analysis of the optical spectroscopy
of J1628, updating the preliminary results presented in Torres et
al. (2004a). The paper is structured as follows: Section 2 presents the
observations and the data reduction procedure. In Section 3 we
estimate the orbital ephemeris and in Section 4 discuss the spectrum
of the secondary star. Note that hereafter, we will call the visible
star in the binary the secondary and the invisible stellar component
the primary, in analogy with the terminology used in the study of
binaries harboring compact objects. In Section 5 we derive an upper
limit to the extinction toward J1628. Section 6
describes the emission lines observed in the spectrum. Finally, in
Section 7 we discuss our results and a summary is given in Section 8.

\section{OBSERVATIONS AND DATA REDUCTION}

The observations of J1628 were obtained using different spectrographs
mounted on the 6.5m Magellan Baade and Clay telescopes at LCO.

The Boller and Chivens (B\&C) spectrograph was used to acquire low
resolution spectra. A 0.7 arcsec slit width and a 600 (1200) line
mm$^{-1}$ grating yielded a spectral resolution of 2 pixels
and a dispersion of 1.56 (0.80)~\AA~pix$^{-1}$. The spectral interval
covered by the B\&C was set depending on the requirements of the
observer's scientific program and can be found in Table 1.

The Inamori-Magellan Areal Camera and Spectrograph (IMACS; Bigelow \&
Dressler 2003) was employed in short-camera mode to obtain
intermediate resolution spectra dispersed along the long-axis of two
of the 8 SITe CCDs in the IMACS detector. Using a 600 line
mm$^{-1}$ grism and a 0.5 arcsec slit width yielded a dispersion of
0.48 \AA~pix$^{-1}$ (CCD \#2) and 0.58 \AA~pix$^{-1}$ (CCD \#5) in the
spectral intervals 5670-7620~\AA~and 7695-10025~\AA~respectively. The
spectral resolution was $\sim2$~pixels FWHM.

High-resolution spectra were acquired with the Magellan Inamori
Kyocera Echelle (MIKE; Bernstein et al. 2003). During May 2002, data
were obtained using a 0.7 arcsec slit and the CCD detector binned by 2
in the spatial direction. In the blue, the useful wavelength range
covered from 3360 to 4700~\AA~over 32 echelle orders. The dispersion
was of 0.017-0.023 \AA~pix$^{-1}$ and the spectral resolution
$\sim5.4$ pixels FWHM. In the red, the useful wavelength range covered
the 4775-8500~\AA~interval over 31 echelle orders with a dispersion
varying between 0.034 and 0.060 \AA~pix$^{-1}$. The spectral
resolution was $\sim4.5$~pixels FWHM. In July 2004 a new dichroic
and CCD were available. High-resolution spectra were obtained using
again a 0.7 arcsec slit, but this time the CCD detector was binned by
2 both in the spatial and spectral direction. With this configuration
the wavelength interval 3325-5070 \AA~was covered over 35 orders and
the interval 4705-7260~\AA~over 25 orders. The 0.7 arcsec slit yielded
a dispersion of 0.033-0.050 \AA~pix$^{-1}$ and a spectral resolution
of 2.7 pixels FWHM in the blue and 0.066-0.10 \AA~pix$^{-1}$ and 2.2
pixels FWHM in the red.

Table 1 provides a detailed journal of the J1628 observations. In
addition to the spectra of J1628, spectra of several radial velocity
standards, flux standards and other stars were acquired during the
observations.

The B\&C and IMACS images were bias and flat-field corrected with
standard {\sc iraf}\footnote{{\sc iraf} is distributed by the National
Optical Astronomy Observatories.} routines. The spectra were extracted
from each CCD frame with the {\sc iraf kpnoslit} package. The
pixel-to-wavelength calibration was derived from cubic spline fits to
HeNe or HeNeAr arc lines. The root-mean square deviation of the fit
was $\leq0.07$~\AA~and $<0.02$~\AA~for the data acquired with the B\&C
and IMACS respectively. Checks for the stability of the wavelength
calibration were made using the strongest atmospheric emission lines
present in the spectrum. For the spectra acquired with B\&C we made
use of the [{\sc O i}] 5577.34 \AA~line and estimated an accuracy in the
wavelength calibration $\leq 0.30$~\AA~(600 line mm$^{-1}$ grating)
and $<0.08$~\AA~(1200 line mm$^{-1}$ grating). In the case of IMACS,
we made use of the [{\sc O i}] 6300.3 \AA~line and the OH emission blend at
7316.3~\AA~(Osterbrock et al. 1996). The accuracy estimated in this
way was $\leq0.08$~\AA~for the wavelength calibration of the
spectra recorded in CCD \#2.

The MIKE data were processed with {\sc iraf} and the spectra extracted
with the {\sc iraf echelle} package and the aid of {\sc iraf} tasks
developed and kindly provided by Professor Jack Baldwin. A dispersion
solution was derived from a two-dimensional fit to the ThAr comparison
lamp spectra. More than 400 lines were included in the fit, and
residuals were typically $<0.0012$~\AA~in the blue and
$<0.0023$~\AA~in the red. No corrections for terrestrial lines were
applied to the data. The atmospheric NaD emission at
$\lambda\lambda5889.95, 5895.92$ and the atmospheric O$_2$ bands with
wavelengths provided in Pierce \& Breckinridge (1973) were used to
establish an accuracy of the fit $<0.02$~\AA.

\section{RADIAL VELOCITY MEASUREMENTS}

The first task was to measure the radial velocities of the optical
counterpart. These were measured from the spectra by the method of
cross-correlation with a template star (Tonry \& Davis 1979). Because
no common template star was observed with all setups, we decided to
use for the cross-correlation the spectrum of a template star from the
Indo-US library of Coud\'e feed stellar spectra (Valdes et
al. 2004). This library contains spectra of 1273 stars at a dispersion
of $0.4$~\AA~pix$^{-1}$ and 1~\AA~FWHM resolution. On basis of the
preliminary results for the spectral class of J1628 (Torres et
al. 2004a), we selected the spectrum of a K3III star (HD169191) for the
cross-correlation. Prior to the cross-correlation, the B\&C and IMACS
spectra were resampled onto a logarithmic wavelength scale and
normalized by dividing with the result of fitting a low order spline
to the continuum. The archival K3III spectrum was resampled onto the
same logarithmic wavelength scale as for the B\&C and IMACS target
spectra, broadened to match the IMACS/B\&C spectral resolution using a
Gaussian function with the appropriate width and normalized to the
continuum. In the case of the MIKE spectra, we chose two orders
covering the wavelength intervals $\lambda\lambda6055-6225$ and
$\lambda\lambda6400-6580$ to measure the radial velocities. These
spectral ranges contain a number of moderately strong lines and blends
useful not only for the radial velocity measurements, but also for the
spectral-type/luminosity classification and rotational broadening
measurement (see next section). Both orders were rebinned to match the
template logarithmic wavelength scale and rectified to the continuum
as explained above.
   
Individual velocities were extracted by cross-correlation with the
archival template star in the range $\lambda\lambda4900-5560$ (B\&C;
Oct 2002), $\lambda\lambda4990-6540$ after masking the interstellar
NaD lines (B\&C; 12 May 2003), $\lambda\lambda6400-6545$ (IMACS) and
after masking the H$\alpha$ emission line in the MIKE data. To obtain
an estimation of the systematic errors caused by the use of the
archival template we proceeded as follows: first, we obtained the
radial velocities for each night (when more than one target spectrum
was available) using now a template spectrum observed during the
respective night. Next, we calculated for each night the radial
velocity differences and compared the values (night per night) with
the radial velocity differences obtained using the archival
template. We found in this way rms errors of 2 km s$^{-1}$ and a
maximum deviation of 5 km s$^{-1}$.

We performed least-squares sine fits to our radial velocity data using
the photometric period P$_{ph}$=4.9364~d and 2$\times$P$_{ph}$ as
initial guesses for the orbital period. A spectroscopic period of
4.93956 $\pm$ 0.00025~d provided the lowest value of the $\chi^2$ per
degree of freedom, being 5 times lower than that obtained for
2$\times$P$_{ph}$. This value is consistent with the deepest minimum
at $\sim0.2$ cycle~d$^{-1}$ observed in the $\chi{^2}$-periodogram of
the radial velocities. Furthermore, this period agrees with the
photometric period within 1.8-$\sigma$. Nevertheless, the  orbital
period is not determined unambiguously because the spectroscopic data
set is also consistent with several alias periods. In this work we adopt
the following parameters of the radial velocity curve:

\begin{itemize}

\item[] $K_{2} = 33.3 \pm 0.6$ km s$^{-1}$

\item[] $\gamma = 14.7 \pm 0.6$ km s$^{-1}$

\item[] P$_{sp}$=4.93956 $\pm$ 0.00025~d

\item[] $T_{0} = HJD~2452575.11 \pm 0.03$

\end{itemize}

The zero phase is defined as the time of closest approach of the
secondary to the observer. All quoted uncertainties are 1-$\sigma$ and
were obtained after increasing the error in the radial velocities in
order to give $\chi^2_{\nu}=1$. The phase-folded radial velocity curve
is shown in Figure 1. Orbital phases of the observations, radial
velocity measurements with their associated statistical errors and
residuals to the circular orbit solution are given in Table
2. Adopting the values of $K_{2}$ and the orbital period, a mass
function of $f(M{_1})=0.019\pm0.001$~M$_\odot$ is obtained.

A 4.9~d orbital period rules out an ellipsoidal modulation of the
optical light curve (Buxton et al. 2004) and suggests a
chromospherically active binary scenario for J1628 where an evolved
stellar component is covered with starspots that modulate the
photospheric light with stellar rotation. The 0.30 mag modulation
observed in the V-band light curve is at the high end for
chromospherically active binaries: only thirteen of 206 binaries
listed in Strassmeier et al. (1993) have shown a modulation in the
V-band with an amplitude $\geq0.3$~mag.

\section{Spectral Type, Luminosity Class and Rotational Broadening}

In a preliminary report based on the analysis of the first B\&C and
MIKE data sets (Torres et al. 2004a), the spectra of J1628 were
compared visually with the spectra of K subgiants and main sequence
stars acquired during the observations and also compared with the
spectra of G and K-type V/IV/III stars from the Indo-Us Library of
Coud\'e feed stellar spectra. Based on the depth of the TiO bands at
$\lambda\lambda 6080-6390$, a spectral type in the range K3$\pm$1 was
supported. A luminosity class III-IV was suggested because of the
strength of the Ca{\sc i}~$\lambda6450$ line with respect to the
metallic lines in the interval~$\lambda\lambda6420-6530$ and by
comparing the relative intensities of the multiplet 33 Ti{\sc i}
lines~$\lambda\lambda8379,8382$ and Fe{\sc i}~$\lambda8388$.

To verify our visual classification and determine the rotational
broadening of the secondary star, we made a more detailed analysis of
the spectra acquired with MIKE during 2004 as they have a higher
signal-to-noise in the red than the 2003 MIKE data due to the binning
in the dispersion direction and the use of a new dichroic. We focused
our analysis in the wavelength interval $\lambda\lambda6350-6530$
where there are several temperature and gravity sensitive lines for F,
G and K stars (Strassmeier \& Fekel 1990; Strassmeier \& Schordan
2000). We used the technique outlined in Marsh, Robinson \& Wood
(1994) that is based on the search of the lowest residual obtained
when  subtracting a set of templates from the Doppler-corrected
average spectrum of the target. The template spectra are broadened
prior to subtraction to determine the rotational broadening of the
lines. This technique also allows the possibility of a continuum
contribution from an accretion flow when searching for the parameters
of the secondary star. To implement this procedure we used a set of
K-dwarf and subgiant template spectra taken with MIKE during 2004 and
a set of K-giant spectra observed with MIKE in March 2003 (see Martini
\& Ho 2004).

We proceeded as follows: first, the target spectra were
Doppler-corrected to the rest frame of the secondary star by
subtracting the radial velocity obtained from the cross-correlation
with the template. Next, we produced an average spectrum after
assigning different weights to the individual spectra in order to
maximize the signal-to-noise of the sum. The template spectra were
then broadened from 37 to 47 km s$^{-1}$ in steps of 0.05 km s$^{-1}$
through convolution with the rotational profile of Gray (1992). We
adopted a linearized limb-darkening coefficient of $0.65$. Each
broadened version of the template spectrum was multiplied by a factor
$f$ (representing the fractional contribution of light from the
secondary star) and subtracted from the target Doppler-corrected
average. Then a $\chi^2$ test on the residuals was performed in the
range $\lambda\lambda6400-6548$ and the optimal values of $f$ and $v
\sin i$ were provided by minimizing $\chi^{2}$. The results from the
$\chi^2_\nu$ minimization are listed in Table~3, with quoted
uncertainties corresponding to $\chi^{2}_{\rm min}$+1 (Lampton, Margon
\& Bowyer 1976). The minimization of $\chi^2_\nu$ in the V/IV
luminosity class templates shows that the spectral type of the
secondary star in J1628 is most likely not later than
K3. Additionally, for a single-lined chromospherically active binary
star the fractional contribution of the secondary must be $\lesssim
1.0$. This rules out a main sequence star and constrains the spectral
type to be K3IV/III. The high $f$ value as determined from the
{${\chi}^2$} minimization using the IV/III templates together with the
fact that the primary star is not detected in the nightly averaged
spectra (which have signal-to-noise of about 70) suggest that the
visual luminosity of the primary star and/or the accretion flow can be
at most a few per cent of the flux the secondary star.

A rotational broadening measurement ($v \sin i$) of 43 km s$^{-1}$ was
obtained from the K3III template. To check the systematic errors
introduced by the choice of the linearized limb-darkening coefficient
(which is only suitable for the continuum; Collins \& Truax 1995), we
have allowed it to vary in the range 0.0-1.0. This leads to 7 per cent
changes in the resulting value of $v \sin i$ (i.e. $\sim$ 3 km
s$^{-1}$) and $f$.  Therefore it is the uncertainty in the
limb-darkening coefficient and not the statistical noise which limits
our accuracy. We therefore adopt a value of $v \sin i = 43 \pm 3$~km
s$^{-1}$, which encompasses all $v \sin i$ values obtained for the
templates in Table 3. Figure 2 shows a comparison of the spectra over
the range $\lambda\lambda6400-6475$. It is clear that there is some
excess absorption in the Ca{\sc i} lines (in particular Ca{\sc
i}~$\lambda6439.1$). This could be caused by differences in the
spectral luminosity or/and metallicity between template and target
spectrum.

\section{Interstellar Extinction Upper limit}

An estimate of the upper limit to the color excess of 0.78 mag is
derived from the weighted average H{\sc i} column within one degree
along the line of sight to J1628
($N{_H}=4.51\times10{^{21}}$~cm$^{-2}$; Dickey \& Lockman 1990) and
the relation between $N{_H}$ and $E(B-V)$ of Bohlin, Savage \& Drake
(1978). For this reddening we should expect the $\lambda\lambda6196,
6203$ diffuse interstellar bands (DIBs) to have EWs of about 78 and
220 m\AA~ respectively (Herbig 1975). However, the EWs of the
absorption features close to these wavelength positions are about 10
m\AA~indicating that the extinction towards J1628 is significantly
lower.

We have searched all medium (IMACS) and high-resolution spectra (MIKE)
for other DIBs and atomic interstellar lines. The only interstellar
features we were able to identify unambiguously in the forest of
absorption lines originating in the secondary were the NaD
$\lambda\lambda 5889.95,5895.92$~and K{\sc i} $\lambda7699$ lines
blended with the namesake broader photospheric lines (see Figure
3). The profile of these interstellar lines show a single
component. Taking into account that double or multiple components in
the profile of the K{\sc i} $\lambda7699$ line appear when its EW is
$\gtrsim0.15$~\AA~(Munari \& Zwitter 1997; MZ97), the upper limit to
E(B-V) can be reduced to $\sim0.6$~mag according to the relation
between reddening and EW for this interstellar line (MZ97). We have
measured an EW of $0.08 \pm 0.01$~\AA~for the K{\sc i} interstellar
line after fitting the K{\sc i} blend with a two-Gaussian model (with
one Gaussian to account for the photospheric absorption line and the
other to account for the interstellar component). Using the
calibration of MZ97 we derived an interstellar extinction of
$E(B-V)=0.30\pm0.04$. For this reddening, the $\lambda\lambda6196,
6203$ DIBs should be stronger than observed (Herbig 1975, see also
Figure 5 in Jenniskens \& D\'esert 1994). This discrepancy suggests
that our fit is overestimating the EW for the interstellar line. Given
the unreliable utility of Herbig's calibrations for low reddening and
our uncertain fit to the K{\sc i} blend, we conservatively adopt for
the remaining of this paper $0.0 < E(B-V)\lesssim 0.6$.

\section{Emission Lines in the Spectrum of J1628: Chromospheric Activity Indicators}

Observational evidence of a chromosphere in the visible spectrum
of active stars relies commonly on the existence of emission in the
cores of the Ca{\sc ii} H \& K lines. Apart from this hallmark active
stars can reveal filling-in or strong emission lines in photospheric
lines like H{\sc i} Balmer lines and the Ca{\sc ii} infrared triplet
(see e.g. Linsky 1980, Thatcher \& Robinson 1993 and references
therein). In this regard, there is clear manifestation of stellar
activity in the spectra of J1628:

The Ca{\sc ii} H \& K lines have emission cores with absorption
reversal at the top of the emission that give them a double-peak shape
(see Figure 4). Both Ca{\sc ii} H \& K emission profiles exhibit
variations in the strength of the violet (V) and red-ward (R) peaks,
the violet peak being stronger than the red-ward peak ($V/R>1$) except
on 2004 Jun 9 when the red peak becomes stronger ($V/R<1$). Single K
dwarfs and giants hotter than spectral type K3 commonly show $V/R>1$
asymmetries as observed in the integrated disk of the Sun. This is
often interpreted to be an indication that they have coronae and
chromospheres dynamically similar to the Sun. The $V/R$ ratios $<1$
observed in giants cooler than K4 are considered to be related to mass
outflow in the chromosphere (Stencel 1978). The temporal
variability of the Ca{\sc ii} peak asymmetries is well documented for
Arcturus ($\alpha$ Boo; K2III). Arcturus has shown transitions from
$V/R>1$ to $V/R<1$ in the Ca {\sc ii} K emission core ratio (Chiu et
al. 1977, Gray 1980). Temporal variations of the Ca {\sc ii} emission
cores have been observed in  other late-type stars and are described
in Rebolo et al. (1989), Garc\'\i a L\'opez et al. (1992) and
references therein. In the case of binary systems, Baliunas \& Dupree
(1982) found that the strength of the Ca {\sc ii} emission profile for
the single-lined (G8III-IV) chromospherically active binary
$\lambda$~Andromedae increases (decreases) at the time of continuum
light minimum (maximum) which correspond to the time when spotted
(unspotted) regions dominate the stellar disk. Moreover, they found
$V/R < 1$ only at maximum light and $V/R > 1$ at other phases and
suggested an explanation of the variations in the profile asymmetry as
due to differential downward and upward motions in the stellar
atmosphere. Unfortunately our observations of J1628 are insufficient
to corroborate the above correlation and explanation. Clearly,
high-resolution spectroscopy with a better sampling of an orbital
cycle and a simultaneous light curve are required.

The mean emission line width ${W_0}(K)$ measured for the Ca {\sc ii}
K line in the MIKE spectra is 0.82~\AA, which yields ${W_0}(K)= 62.5 $~km
s$^{-1}$ after the quadratic correction of the instrumental broadening
(18.3 km s$^{-1}$). From ${W_0}(K)$, we estimated an absolute visual
magnitude of 2.0 for the secondary of J1628 by using the Wilson-Bappu
relation for chromospherically active binaries (Montes et al. 1994):
${M_V}=-16.01 \log {W_0}(K) + 30.79$, where ${W_0}(K)$ is expressed in
km s$^{-1}$. However J1628 may deviate significantly from the
Wilson-Bappu law due to the influence of its high rotational
broadening. Montes et al. (1994) found that ${M_V}$ was overestimated
for large values of $v \sin i$, up to 2 mag when $v \sin i \sim 40$~km
s$^{-1}$ (see figure 5 in their paper). Therefore ${M_V}$ could be
$\sim4.0$ for the secondary in J1628. In any case, these
estimations of ${M_V}$ are in between the values for a K3
dwarf (${M_V}=6.8$; Gray 1992) and a K3 giant (${M_V}=0.3$). This
result is in agreement with the expected evolved secondary for J1628.

The H$\alpha$ line is in emission above the continuum (see Figure 4)
with EW values between 0.3-1.3~\AA, except on night 2003 May 6, when
the EW increases to 4~\AA. This may be due to intrinsic activity
variations, like a flare eruption. The H$\alpha$ profile obtained from
the high resolution spectra has a FWHM$\sim160-250$ km s$^{-1}$ and shows
a self-reversal core during some of the nights. Variable broad
H$\alpha$ emission (see e.g. Byrne et al. 1995
for an insight on the broadening mechanism) sometimes with a
self-reversal core has been observed among the most extreme
chromospherically active binary stars, for instance UZ Lib (K0III,
P$_{sp}$=4.76~d; Bopp et al. 1984), II Peg (K2IV, P$_{sp}$=6.72 d;
Byrne et al. 1995), EZ Peg (G5IV-III/K0V, P$_{sp}$=11.6 d; Montes et
al. 1998), XX Tri (K0III, P$_{sp}$=24~d; Bopp et al. 1993) and HD6139
(K2IV-III, P$_{ph}$=31.95 d; Padmakar et al. 2000). The EW of the
observed H$\alpha$ emission line measured in these systems is of the
order of 1~\AA~(see the above references). For comparison, the
H$\alpha$ EWs measured in X-ray novae in quiescence are of the order
of tens to hundreds of Angstroms. For instance, V404 Cyg (K0IV,
P$_{sp}$=6.47~d, $v \sin i = 38$ km s$^{-1}$) and Cen X-4 (K3-5V,
P$_{sp}$=0.63~d, $v \sin i = 43$ km s$^{-1}$) have respectively
H$\alpha$ EWs of 38~\AA~(Casares et al. 1993) and 35~\AA~(Torres et
al. 2002).

\section{DISCUSSION}

Tidal theory (Zahn 1977) predicts that late-type stars in close binary
systems rotate in synchronization with the orbital motion because
tidal interactions are effective in forcing synchronization on time
scales shorter than the evolutionary life time of the systems. A
purely hydrodynamical mechanism based on the effects of meridional
currents in the atmospheres of the stellar components and due to their
non-spherical shape also explains the synchronism observed in close
binary systems (Tassoul \& Tassoul 1992). The estimated spectroscopic
(orbital) and photometric (rotational) periods for J1628 are nearly
identical implying synchronism. The small difference (if real) could
be due to changes in the spot pattern over the years. Using the
measured rotational broadening (section 4) and assuming that the
secondary star is spherical, we can obtain a lower limit for the
radius of the secondary star: $R{_2} \ge R{_2} \sin i~(R_\odot)=
P{_{rot}}~v \sin i / 2 \pi = \frac{1}{50.6}~P{_{rot}~(days)}~v \sin
i~(km/s)= 4.2~R_\odot$. A K3V star has a $\sim0.7~R_\odot$~radius
(Gray 1992) and a K3III star has a radius of $20.5\pm0.6~R_\odot$
(van Belle et al. 1999). Assuming the secondary is not larger than a
giant, we obtain a minimum value of the binary inclination of $\arcsin
i = R{_2} \sin i / R{_{K3III}} = 12^o$.

When the secondary star in a synchronous binary fills its
Roche lobe, the mass ratio $q={M_2}/{M_1}$ can be determined through the
expression (see e.g. Wade \& Horne 1988 and Eggleton 1983):

$$
v \sin i =
K_{2}~(1 + q) {0.49q^{2/3}\over{0.6q^{2/3}+\ln \left( 1+q^{1/3}\right)}}
$$

From the values of $v \sin i$ and $K_2$ found for J1628, we derive a
lower limit of $q~>~2.0$, which implies a velocity semi-amplitude of
the primary ${K_1}=q{K_2}~>~67$~km s$^{-1}$. These are lower limits
because the radius of the secondary in J1628 may be smaller than its
Roche lobe and the latter increases its size with $q$. Assuming mild
mass exchange/loss during the evolution of the system (Popper \&
Ulrich 1972, Dupree 1986), the mass of the secondary should be in the
range of masses expected for K3V to K3III stars, i.e. 0.7 to
1.1~M$_\odot$ (Lang 1992). Hence ${M_1}={M_2}/q <  0.5 {M_2} \lesssim
0.6$~M$_\odot$. On this basis, the unseen primary of J1628 is probably
a dwarf (if not a white dwarf) of spectral type K7 or later.  A
stringent lower limit for the binary inclination of $i \gtrsim 41^o$
is derived from the mass function
($f(M{_1})=(1+q)^{-2}~{M_1}{{\sin}^3} i$) when using the upper limit
for ${M_1}$, the lower limit for $q$ and the value of $f(M{_1})$. The
radius of the secondary is thereby $\lesssim R{_2} \sin {41^o} / \sin
{41^o}= 6.4~R_\odot$. In short, $4.2~R{_\odot} \le R{_2} \lesssim
6.4~R{_\odot}$ for the visible stellar component in J1628. These
constraints on the radius can be used in conjunction with the
constraint in the spectral type of the secondary (Section 4) to obtain
its absolute visual magnitude by applying Equation 2 of Popper
(1980). Using the visual absolute flux (surface brightness
parameter) for a K3V/III star (Table 1 in Popper 1980) we derive $2.0
< M{_V} < 3.8$ in harmony with the absolute magnitudes estimated from
the Wilson-Bappu relationship (Section 6). From $M{_V}$, the apparent
magnitude $V=13.4$ and the constraints to the reddening obtained in
Section 5 we find 350~pc $ < d < 1.9$~kpc for the distance to J1628.

We made use of PIMMS to convert the ROSAT PSPC count rates to an
unabsorbed flux of $8.5\times10^{-13} < f{_x} \lesssim
4.4\times10^{-12}$~erg cm$^{-2}$ s$^{-1}$ (0.1-2.4 keV) by applying an
absorbed Raymond Smith model with $\log T=7$, assuming a metal
abundance of 0.2 times the solar value and $0.0 <{N_H}\lesssim
3.48\times10^{21}$~cm$^{-2}$ (see Yi et al. 1997 regarding the utility
of using this 1-T model to obtain X-ray fluxes). Furthermore, for
each possible value of ${N_H}$ and $M{_V}$ we can evaluate the
distance towards J1628 using the distance modulus and (as above) the
corresponding intrinsic X-ray flux. This yielded an intrinsic X-ray
luminosity of $6.6\times10^{31} < L{_x} < 6.4\times10^{32}$~erg
s$^{-1}$~\footnote{The range in computed $L{_x}$ is
somewhat less than the square of the range in the distance error
because of the complex interplay between the assumed ${N_H}$ and the
derived distance. High {$N_H$} yield high A${_V}$ and
lower distances (therefore lower computed $L{_x}$), but the X-ray
absorption correction factor is higher at high ${N_H}$ which increases
the computed $L{_x}$ and somewhat counteracts the distance
effect.}, well above the averaged X-ray luminosity observed in
chromospherically active binaries ($\sim1\times10^{30}$~erg s$^{-1}$;
Padmakar et al. 2000), but consistent with the X-ray luminosities for
RS CVn stars ($1\times10^{29.2-32.2}$~erg s$^{-1}$). This range of
X-ray luminosities for RS CVn stars was derived from observations with
Einstein IPC (energy band 0.16-4~keV) and ROSAT PSPC (see Drake et
al. 1989, 1992 and Dempsey et al. 1993). The differences in the X-ray
flux due to the different bandpasses are expected to be $\lesssim
10\%$ (Dempsey et al. 1993; Benz \& G\"{u}del 1994).

An upper limit to the quiescent radio flux density of $0.3$~mJy was
obtained from observations (Slee et al. 2002, Rupen et al. 2004) and
can be used to check if the radio flux is consistent with that
expected for a quiescent chromospherically active binary. We compared
this with the quiescent radio luminosity found for chromospherically active
binaries of $\log L{_{rad}}=(1.37 \pm 0.09) \log L{_x}-26.38$
(Padmakar et al. 2000; see also Drake, Simon \& Linsky 1989 and Benz
\& G\"{u}del 1994). In this way we derived 0.2 $< f{_{rad}} < 2.0$~mjy
for J1628. While this range is consistent with the observed upper
limit, the range is due almost entirely to our uncertainty in the
X-ray luminosity - the correlation between $L{_x}$ and $L{_{rad}}$ is
rather tight. If the true X-ray luminosity is at the high end of our
range, then the quiescent radio flux is inconsistent (the predicted
radio flux is larger than the observed upper limit). In this regard,
the hard X-ray emission observed for J1628 (ROSAT hardness ratio
HR1=1.00) is common for RS CVn stars undergoing an X-ray flare
(see e.g. Graffagnino, Wonnacott \& Schaeidt 1995). The radio
counterpart to J1628 occasionally increases its flux density to
$0.35-13.8$~mJy at 8.6 and 4.8 GHz (Tsarevsky et al. 2001; Rupen et
al. 2002,2004; Slee et al. 2003). Transient radio brightenings with
similar or larger amplitude to those observed in J1628 are frequent
in chromospherically active binary stars (Slee et al. 1987, Drake et
al. 1989).

Finally, we considered the possibility that a foreground object could
be the source of the radio and X-ray emission. We searched the images
obtained with IMACS during the acquisition of the spectra to find a
fainter optical source nearby GSC 07861-01088, the proposed (brighter)
optical counterpart (see Figure 5). We performed an analysis of
optical astrometry from one of the images to check the positional
coincidence of GSC 07861-01088 and its radio counterpart (Rupen
et al. 2004). The agreement between the optical and radio position is
well within the sub-arcsecond astrometric error and we rule out the
possibility that the nearby fainter object, which is located 2 arcsec
South at RA (J2000)=16:28:47.27 and DEC(J2000)=-41:52:41.0 is the
radio source.

Optical spectroscopy obtained for other microquasar candidates
has shown that they are mostly extragalactic in nature and, in the few
stellar cases, likely chromospherically active stars/binaries
(Mart\'\i~et al. 2004a,b; Torres et al. 2004b, Tsarevsky et
al. 2005). Hence the search for microquasars using X-ray and radio
surveys has not provided so far any new and secure microquasars. This
suggests that there are few microquasars with persistent bright
radio/X-ray emission to be found with the RASS Bright Source
catalogue. Apart from the discovery of X-ray transient
microquasars (e.g. Garcia et al. 2003), the progress in
understanding relativistic outflows in X-ray binaries depends largely on
observations of known X-ray binaries using current and future
instruments with higher sensitivity and angular resolution. Suitable
targets are for instance radio emitting X-ray binaries where jets have
not been resolved yet (see e.g. Mirabel \& Rodr\'\i guez 1999, Fender
2004b) and X-ray transients in quiescence where radio jets are
expected at a level of a few $\mu$Jy (see e.g. Gallo, Fender \& Pooley
2003; Gallo, Fender \& Hynes 2005).

\section{CONCLUSIONS}

We have presented comprehensive optical spectroscopy of the
microquasar candidate 1RXS J162848.1-41524. From the analysis of the
absorption line spectrum, we have determined an orbital period of
P$_{sp}$=4.93956 $\pm$ 0.00025~d and a radial velocity semiamplitude
of the secondary of $K_{2} = 33.3 \pm 0.6$ km s$^{-1}$. The implied
mass function is $f(M{_1})=0.019\pm0.001$~M$_\odot$. We have
established the rotational broadening of the secondary star to be $v
\sin i = 43 \pm 3$ km s$^{-1}$. This provides an upper limit to the
mass ratio of $q>2.0$. A $\chi^2$ test applied to the residuals
obtained by subtracting different template stars to the
high-resolution spectra (Section 4), the limits to the absolute visual
magnitude (Sections 6, 7) and the constraints to the stellar radius
(Section 7) support a K3IV secondary. The emission lines observed in
the spectrum of J1628 are consistent with those observed in
chromospherically active binaries. This fact together with the results
presented above indicates a chromospherically active binary and not a
microquasar nature for J1628, where the X-ray and radio emission
are powered by the stellar chromosphere and the periodic photometric
variability reported by Buxton et al. (2004) would be due to cool
surface spots. No trace of the primary has been found in the
photospheric spectrum leading to the conclusion that it is a low
luminosity object, possibly a late-type K dwarf or white dwarf. Our
analysis is hampered however by the fact that our spectroscopic
observations have not the required temporal coverage to determine
unambiguously the orbital period of J1628. Therefore a better sampled
radial velocity curve is still necessary to confirm the parameters
derived in this paper for J1628. Photometric observations at different
epochs will be enlightening: if J1628 is a chromospherically active
binary system, it should show significant changes in the shape and
amplitude of the light curve due to variations in the starspot
distribution over the surface of the secondary.

\section*{ACKNOWLEDGMENTS}

We thank J. Bloom, M. Holman and S. Laycock for helping during the
observations. We also thank P. Martini for providing some of the MIKE
templates. Use of {\sc molly}, {\sc doppler} and {\sc trailer}
routines developed largely by T. R. Marsh is acknowledged. We
also thank the referee for useful comments. This research has made
use of the SIMBAD database, operated at CDS, Strasbourg, France. This
work was supported by NASA LTSA grant NAG-5-10889 and NASA contract
NAS8-39073 to the Chandra X-ray Center. DS acknowledges a Smithsonian
Astrophysical Observatory Clay Fellowship.

\clearpage
\begin{table}
\begin{center}
\caption{Journal of Observations.}
\begin{tabular}{lccccc}
\tableline\tableline
Date              & Instrument   & No.         & Exp. time & $\lambda$ range & Dispersion       \\
(UT)              &              & spectra     &   (s)     &  (\AA)          & (\AA~pix$^{-1}$) \\
\\
26-28 Oct 2002    &  B\&C        & 11          &    300    &   3860-5500     & 0.80              \\
6 May 2003        &  B\&C        & 3           &    120    &   3560-6745     & 1.56              \\
12,13 May 2003    &  MIKE        & 19          &    600    &   3360-4700     & 0.017-0.023       \\
                  &              &             &           &   4775-8500     & 0.034-0.060       \\
13 May 2003       &  B\&C        & 1           &    200    &   5180-6814     & 0.80              \\
22,23,25 Jun 2003 &  B\&C        & 26          &     60    &   4000-7190     & 1.56              \\
17,19,20,22 Jun 2004&  IMACS     & 13          &60,120,$2\times600$ & 5670-7620   & 0.48         \\  
                  &              &             &           & 7695-10025      & 0.58              \\
9-11 Jun 2004     &  MIKE        &  8          & 300,600   & 3325-5070       & 0.033-0.05        \\
                  &              &             &           & 4705-7260       & 0.066-0.1         \\
\tableline
\end{tabular}
\end{center}
\end{table}

\begin{table}
\begin{center}
\caption{Heliocentric radial velocities of J1628.}
\begin{tabular}{cccccc}
\tableline\tableline
\\
& HJD          & $\phi$ & RV           & Error        & O--C \\
& (+2450000.0) &        &(km s$^{-1}$) &(km s$^{-1}$) & (km s$^{-1}$) \\
\\
\tableline
& 2574.4895 & 0.875 & -10.9  &	0.4 &    -2.0  \\
& 2574.4935 & 0.876 & -10.5  &	0.4 &    -1.7  \\
& 2574.4971 & 0.876 & -10.5  &	0.4 &    -1.8  \\
& 2574.5012 & 0.877 & -10.2  &	0.4 &    -1.7  \\
& 2575.4783 & 0.075 &  28.2  &	2.8 &    -1.6  \\
& 2575.4808 & 0.076 &  22.2  &	1.6 &    -7.7  \\
& 2575.4850 & 0.076 &  24.8  &	0.8 &    -5.3  \\
& 2575.4898 & 0.077 &  25.1  &	0.7 &    -5.1  \\
& 2575.4953 & 0.078 &  29.4  &	0.5 &    -1.0  \\
& 2575.4992 & 0.079 &  28.5  &	0.5 &    -2.1  \\
& 2575.5029 & 0.080 &  26.7  &	0.5 &    -4.0  \\
& 2771.6842 & 0.796 & -16.5  &	0.7 &     0.8  \\
& 2771.6898 & 0.797 & -18.7  &	0.6 &    -1.5  \\
& 2771.7011 & 0.800 & -17.4  &	0.4 &    -0.4  \\
& 2771.7099 & 0.802 & -17.2  &	0.3 &    -0.3  \\
& 2771.7181 & 0.803 & -16.4  &	0.4 &     0.4  \\
& 2771.7824 & 0.816 & -16.7  &	0.2 &    -0.9  \\
& 2771.7908 & 0.818 & -16.2  &	0.2 &    -0.5  \\
& 2771.8005 & 0.820 & -15.4  &	0.2 &     0.0  \\
& 2771.8576 & 0.831 & -15.0  &	0.2 &    -0.6  \\
& 2771.8657 & 0.833 & -14.4  &	0.3 &    -0.2  \\
& 2771.8738 & 0.835 & -13.7  &	0.2 &     0.4  \\
& 2771.8885 & 0.838 & -13.0  &	0.2 &     0.7  \\
& 2771.8971 & 0.840 & -12.4  &	0.2 &     1.1  \\
& 2771.9052 & 0.841 & -13.2  &	0.2 &     0.1  \\
& 2771.9133 & 0.843 & -12.6  &	0.2 &     0.6  \\
& 2772.7181 & 0.006 &  23.3  &	0.5 &     7.4  \\ 
& 2772.7677 & 0.016 &  22.3  &	0.3 &     4.3  \\
& 2772.7758 & 0.017 &  23.4  &	0.5 &     5.1  \\
& 2772.7841 & 0.019 &  21.4  &	0.5 &     2.8  \\
& 3173.7346 & 0.190 &  43.4  &	0.6 &    -2.2  \\
& 3173.7378 & 0.191 &  43.9  &	1.0 &    -1.8  \\
& 3173.7408 & 0.192 &  43.7  &	0.8 &    -2.0  \\
& 3173.7430 & 0.192 &  43.0  &	0.8 &    -2.8  \\
& 3175.5575 & 0.559 &   0.4  &	0.6 &    -2.1  \\
& 3175.5593 & 0.560 &   0.5  &	0.5 &    -2.0  \\
& 3175.5611 & 0.560 &   0.6  &	0.6 &    -1.8  \\
& 3175.5629 & 0.560 &  -2.3  &	0.5 &    -4.6  \\
& 3175.5646 & 0.561 &  -0.2  &	0.6 &    -2.5  \\
& 3176.5215 & 0.754 & -13.2  &	0.3 &     5.5  \\	 
& 3178.5249 & 0.160 &  46.1  &	0.4 &     3.3  \\
& 3178.5281 & 0.161 &  45.9  &	0.4 &     3.1  \\
& 3178.5313 & 0.161 &  46.3  &	0.4 &     3.4  \\
& 3195.4598 & 0.588 &  -2.1  &	0.4 &     0.9  \\
& 3195.4662 & 0.590 &  -2.5  &	0.2 &     0.6  \\
& 3196.6135 & 0.822 & -15.3  &	0.3 &     0.0  \\ 
& 3196.6211 & 0.824 & -13.5  &	0.3 &     1.6  \\
& 3196.6286 & 0.825 & -15.2  &	0.3 &    -0.2  \\
& 3197.4637 & 0.994 &  11.5  &	0.3 &    -2.0  \\ 
& 3197.4713 & 0.996 &  11.2  &	0.3 &    -2.6  \\
& 3197.4789 & 0.997 &  12.2  &	0.2 &    -1.9  \\
\tableline
\end{tabular}
\end{center}
\end{table}

\begin{table}
\begin{center}
\caption{Spectral classification and rotational broadening.}
\begin{tabular}{lcccccc}
\tableline\tableline
Template     &Spectral     &  $v \sin i$      &$\chi^2_{\nu}$   &$f$       \\
 	     &Type	   &  (km s$^{-1}$)   &(d.o.f.=1645)    &           \\ 
\\
 
HD88284   &    K0~III	   & 42.5 $\pm$ 0.2  & 2.58  	      & 1.14 $\pm$ 0.01\\ 
HD95272   &    K1~III	   & 43.0 $\pm$ 0.2  & 2.64         & 1.14 $\pm$ 0.01\\
HD43827   &    K3~III$^a$  & 42.8 $\pm$ 0.2  & 2.46  	      & 1.01 $\pm$ 0.01\\ 
          &                                    &              &                \\
HD217880   &    G8~IV	   & 42.7 $\pm$ 0.3  & 3.29  	      & 1.50 $\pm$ 0.02\\ 
HD215784   &    K1~IV	   & 42.9 $\pm$ 0.2  & 2.41         & 1.10 $\pm$ 0.01\\
HD163197   &    K4~IV	   & 43.4 $\pm$ 0.3  & 3.51         & 0.91 $\pm$ 0.01\\
           &                                   &              &                \\
HD223282   &    K0~V	   & 43.2 $\pm$ 0.3  & 3.78         & 1.71 $\pm$ 0.02\\  
HD223121   &    K1~V	   & 42.4 $\pm$ 0.2  & 2.83         & 1.18 $\pm$ 0.01\\       
HD218279   &    K2~V	   & 42.5 $\pm$ 0.2  & 2.64         & 1.18 $\pm$ 0.01\\
HD217580   &    K4~V	   & 42.1 $\pm$ 0.2  & 2.72         & 1.15 $\pm$ 0.01\\
HD130992   &    K5~V$^a$   & 42.4 $\pm$ 0.3  & 3.28         & 1.10 $\pm$ 0.01\\ 
\tableline
\end{tabular}
\begin{tabular}{c}
\\
$^a$ Spectral types adopted from Jasniewick et al. (1999) and Robinson \& Cram (1990).
\end{tabular}
\end{center}
\end{table}

\clearpage
\begin{figure}
\begin{center}
\includegraphics[angle=0,width=5in]{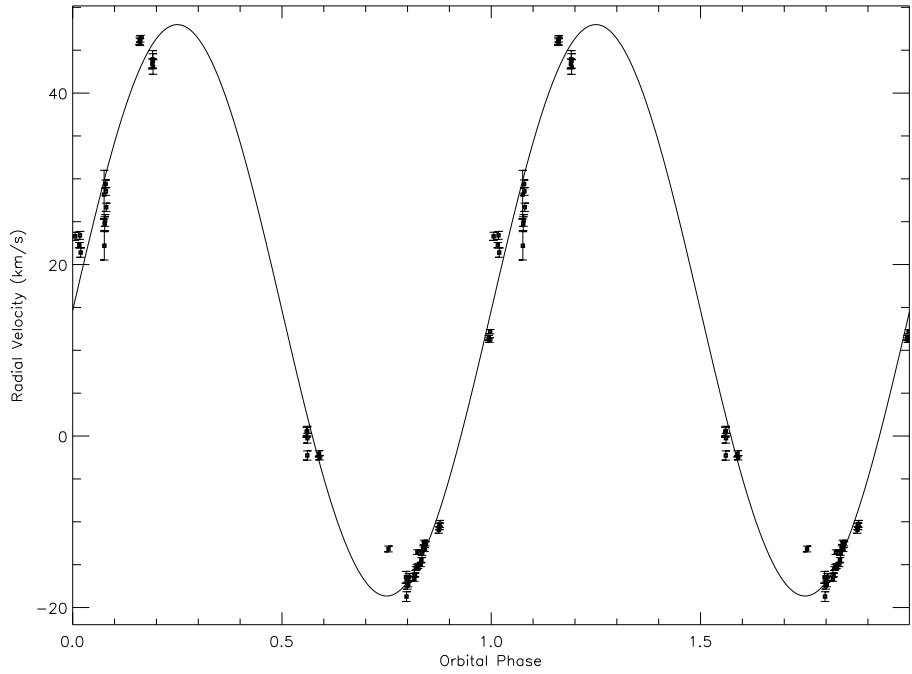} 
\caption[]{Radial velocities of the secondary star in J1628 folded on the ephemeris of Sect. 3. The best sine-wave fit is also shown.}
\end{center}
\end{figure}

\begin{figure}
\begin{center}
\includegraphics[width=5in]{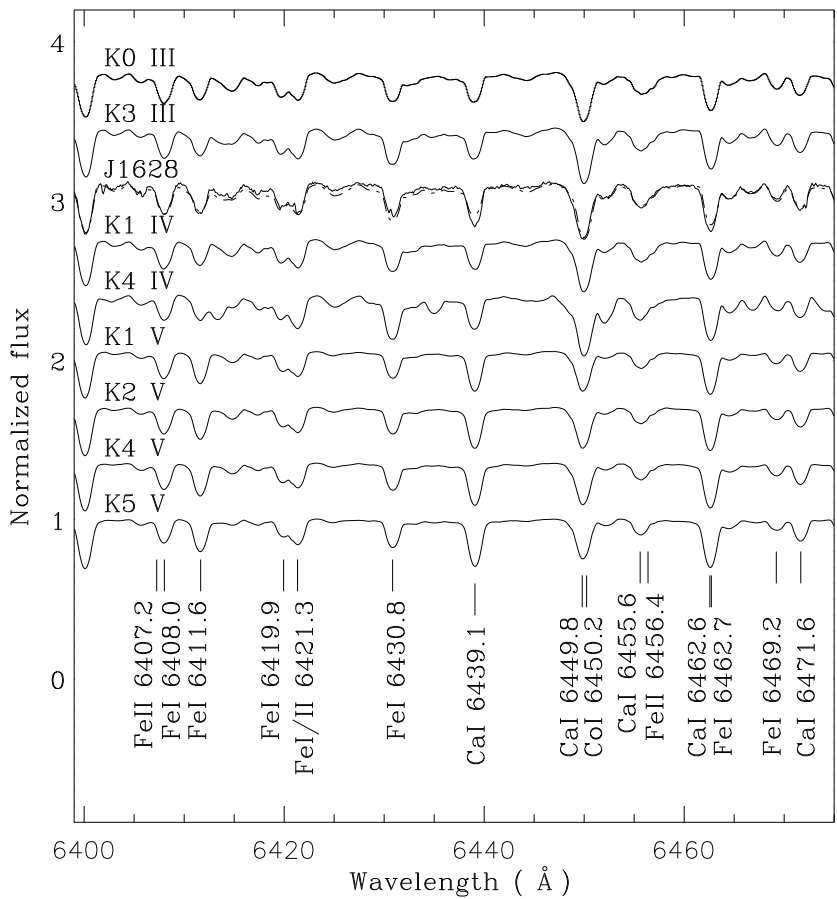}
\caption[]{Averaged spectrum of J1628 and the template spectra after being artificially spun up using the rotational broadenings listed in Table 3. The spectra have been vertically shifted by a constant value for the sake of clarity. The spectrum of HD43827 (K3III) has been superimposed to that of J1628 (dashed line).}
\end{center}
\end{figure}

\begin{figure}
\begin{center}
\includegraphics[width=5in]{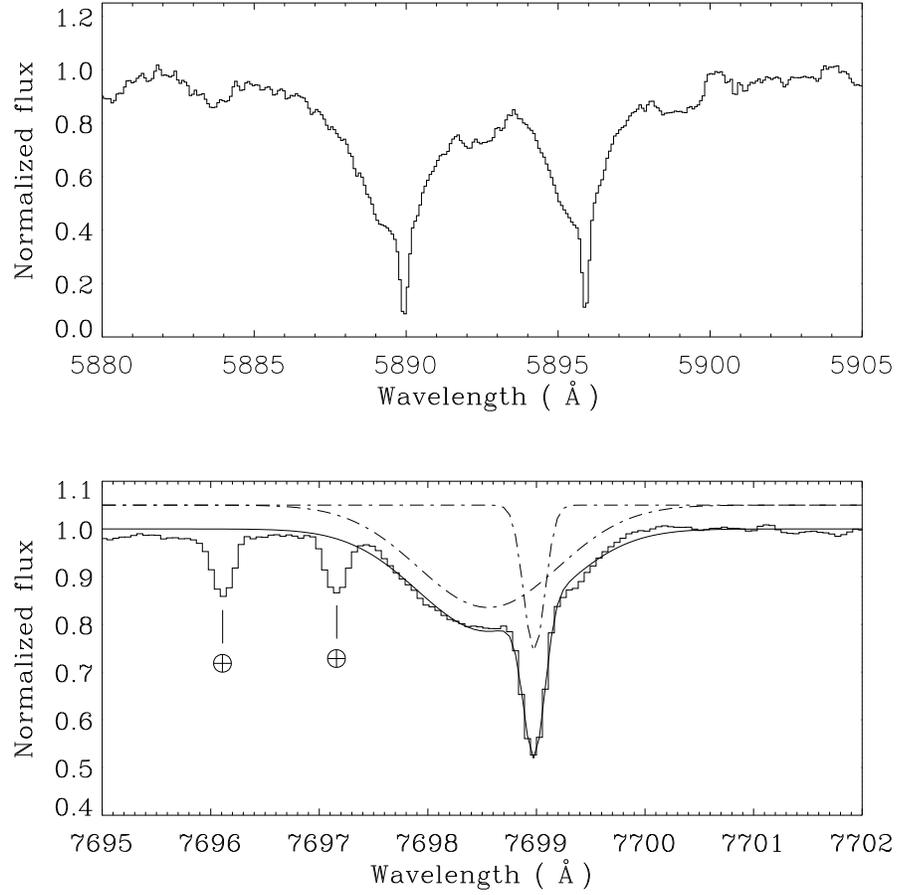}
\caption{Top: the Na D1 $\lambda5889.95$ and Na D2 $\lambda
5895.92$ interstellar/stellar blends. Bottom: K{\sc i} $\lambda7699$ interstellar/stellar blend, showing the two Gaussians fit (dashed line) and their sum. Telluric features are marked with Earth symbols.}
\end{center}
\end{figure}

\begin{figure}
\begin{center}
\includegraphics[width=6in]{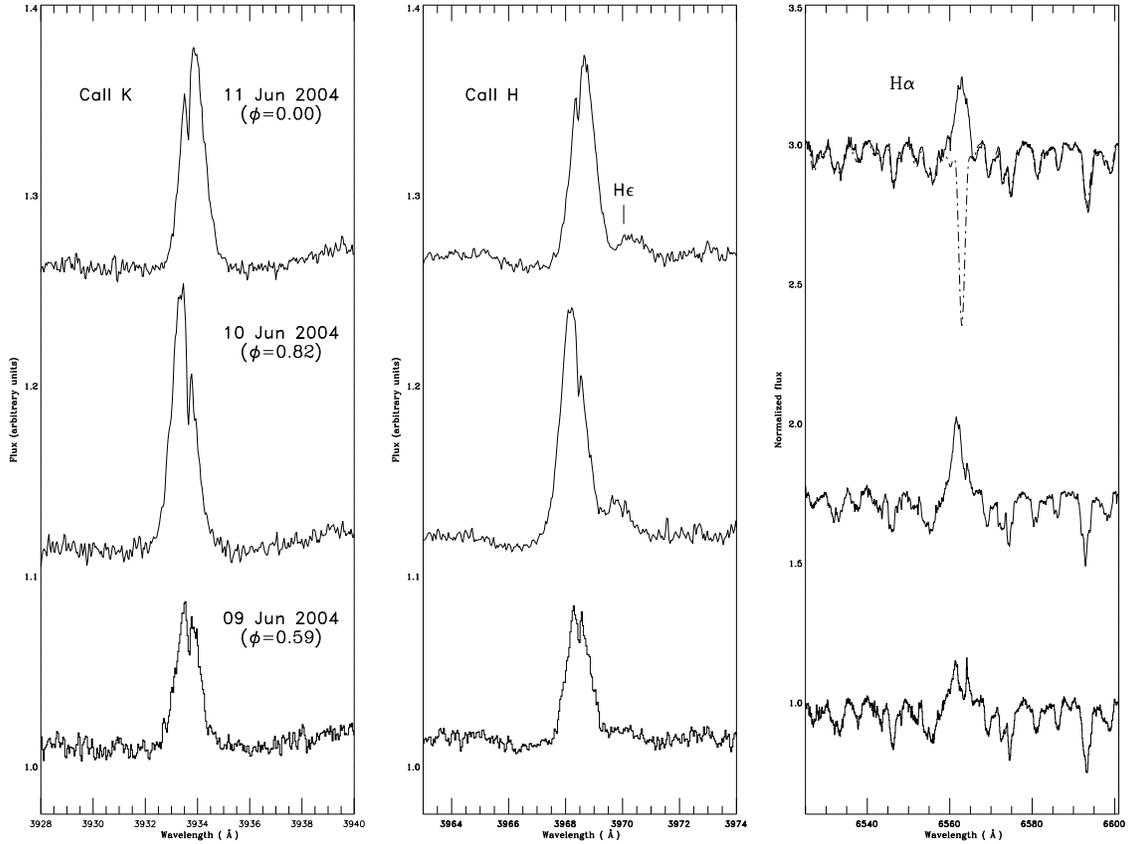}
\caption{The observed night-averaged 2004 MIKE spectra of J1628 in the Ca{\sc ii} and H$\alpha$ regions from Jun 9 (bottom) to Jun 11 (top). The spectrum of HD43827 (K3III) has been superimposed to that of J1628 in the H$\alpha$ region after being broadened to the J1628's rotational broadening (dashed line).}
\end{center}
\end{figure}

\begin{figure}
\begin{center}
\includegraphics[width=6in]{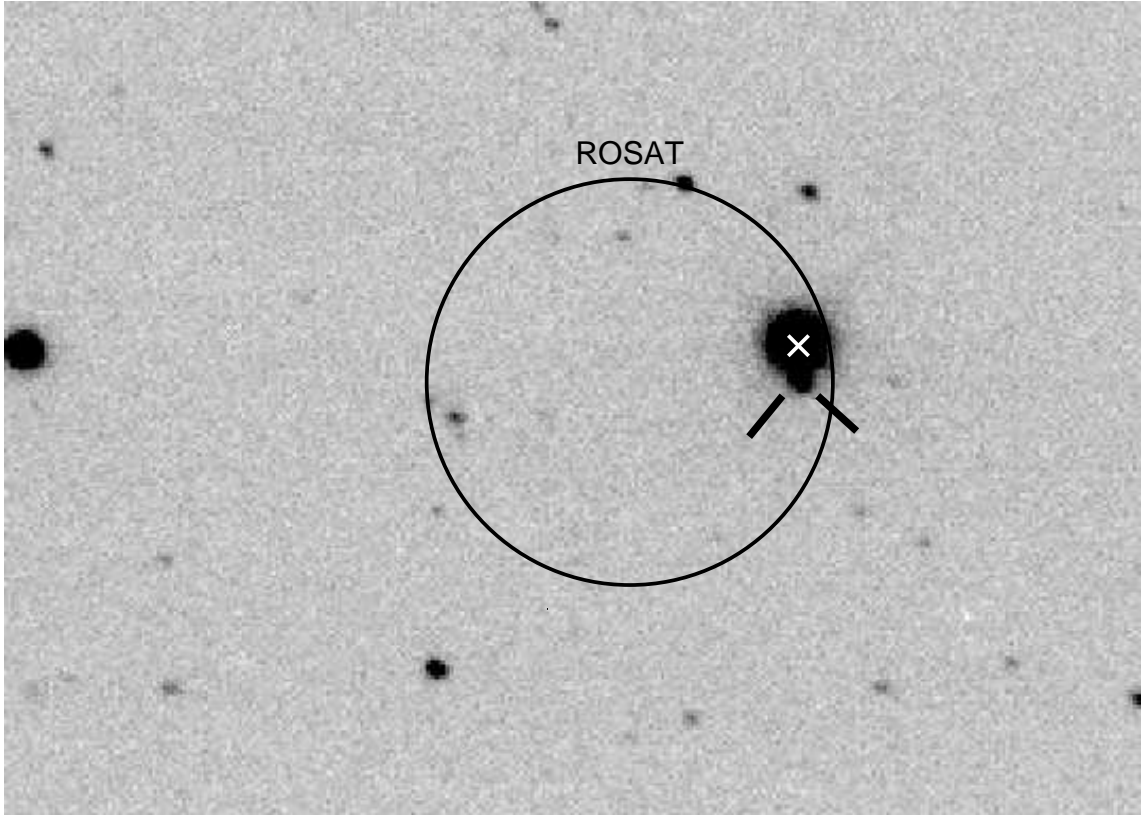}
\caption{One-second R-band image around J1628. North is up. East is to the left. The frame was obtained with a seeing of 0.7 arcsec and a projected pixel size of 0.2 arcsec. The brightest star in the image is GSC
07861-01088, the optical counterpart to J1628. The white cross marks the position of the radio counterpart to J1628. The error circle for the position provided in the RASS Bright Source Catalogue has a radius of 11 arcsec. The linear scale is adjusted to highlight the nearby faint field star located South of the optical/radio counterpart to J1628. Two perpendicular lines point to its position.}
\end{center}
\end{figure}

\end{document}